\documentclass[aps,amsmath,amssymb,twocolumn]{revtex4} 

\usepackage{graphicx}
\usepackage{dcolumn}
\usepackage{bm}

\begin{document}

\bibliographystyle{apsrev} 

\title{All electrical coherent control of the magnetization in thin Yittrium Iron Garnet film}

\author{O. Wid$^{1}$}%
\author{M. Wahler$^{1}$}%
\author{N. Homonnay$^{1}$}%
\author{T. Richter$^{1}$}%
\author{G. Schmidt*$^{1,2}$}%

\affiliation{%
$^1$Institut f\"{u}r Physik, Martin-Luther-Universit\"{a}t Halle-Wittenberg, D-06099, Germany* \\
$^2$IZM, Martin-Luther-Universit\"{a}t Halle-Wittenberg, D-06099, Germany* \\
}

\begin{abstract}
We demonstrate coherent control of time domain ferromagnetic resonance by all electrical excitation and detection. Using two ultrashort magnetic field steps with variable time delay we control the induction decay in yttrium iron garnet (YIG). By setting suitable delay times between the two steps the precession of the magnetization can either be enhanced or completely stopped. The method allows for a determination of the precession frequency within a few precession periods and with an accuracy much higher than can be achieved using fast fourier transformation. Moreover it holds the promise to massively increase precession amplitudes in pulsed inductive microwave magnetometry (PIMM) using low amplitude finite pulse trains. Our experiments are supported by micromagnetic simulations which nicely confirm the experimental results.

\end{abstract}

\maketitle
Coherent control of spin precession is well known from semiconductor spintronics. There for example ultrashort pulses of circularly polarized light are used to induce a spin polarized carrier population in a non-magnetic semiconductor. The spin polarized carriers precess in an external magnetic field and the precession is detected by time resolved Faraday rotation \cite{Crooker1996} or Kerr rotation \cite{Kikkawa1997}. When the excitation pulses are applied in sequence synchronized with the precession frequency the polarization increases or decreases depending on the relative phase of precession and excitation \cite{Yamaguchi2010}, \cite{Hansteen2006}, \cite{Wolf_Diploma_2007}. If a train consisting of a large number of pulses is used to increase the signal far beyond the single excitation case the effect is also named resonant spin amplification \cite{Kikkawa1998}, \cite{Malajovich2000}. While these methods employ optical excitation and/or detection a similar scheme can be envisaged for time domain ferromagnetic resonance using electrical excitation by ultrashort magnetic field steps and electrical detection by induction. As will be shown a main advantage of this resonant excitation especially in PIMM is the massively enhanced resolution. In addition using a synchronized pulse train may reduce the problems which normally arise in PIMM from the unfavorable ratio of the very large excitation and the very small inductive response which are difficult to separate.

\begin{figure}[ht]
\includegraphics[width=8cm]{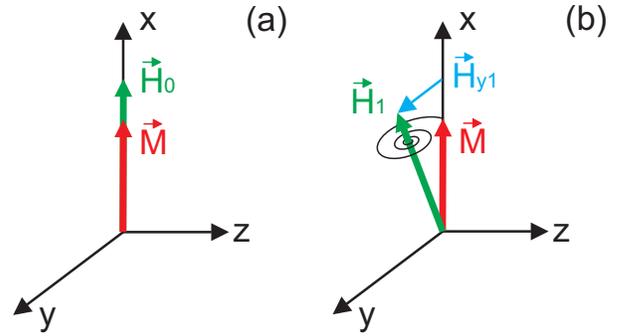}
\caption{Basic concept of Pulsed Inductive Microwave Magnetometry (PIMM). As the initial state the magnetization is aligned along the magnetic field $\vec{H}_{0}$ (a). The first field-step $\vec{H}_{y1}$ tilts the direction of the effective magnetic field, which leads to damped precession of the magnetization around the new direction $\vec{H}_{1}$.
\label{fig:amplification_theory}}
\end{figure}

The basic principle of all-electrical coherent magnetization control is illustrated in figure \ref{fig:amplification_theory}.
We consider a magnetic sample with the magnetization $\vec{M}$ which is aligned along a constant magnetic field $\vec{H}_{0}$ along the x-axis (Fig. \ref{fig:amplification_theory}, (a)). To excite the magnetization out of its equilibrium state we apply a small magnetic field step $\vec{H}_{y1}$ (with $|\vec{H}_{y1}|\ll |\vec{H}_{0}|$) at time $t=0$ with ultra-short rise time along the y-axis, which tilts the magnetic field towards $\vec{H}_{1}$. The angle between $\vec{H}_{1}$ and $\vec{M}$ is dubbed $\Theta$ and at time $t=0$ is identical to the angle between the magnetic field vectors $\vec{H}_{0}$ and $\vec{H}_{1}$ which we name $\Theta_0$. The magnetization now starts to precess around $\vec{H}_{1}$ according to the LLG equation (Fig. \ref{fig:amplification_theory}, (b)) with a precession frequency which is in good approximation $\omega=|\vec{H}_0| \gamma$. The precession amplitude depends on the magnitude of $\vec{H}_{y1}$ or more precise on the angle $\Theta$. This first part corresponds to the basic principle of the PIMM experiment \cite{Silva1999}, \cite{Kos2002}, \cite{Neudecker2006}. To describe the precession in an analytical way we use the approximation of a small precession angle assuming that the change of the x-component of $\vec{M}$ (${M}_x$) is negligible or at least small compared to the maximum change of the components ${M}_y$ and ${M}_z$. We then have
\begin{align}
	{M}_y={M}_0 [1-e^{-t/\tau} \cos{\omega t}],\\
	{M}_z=-{M}_0 [e^{-t/\tau} \sin{\omega t}]
\end{align} with
\begin{align}
	{M}_0=|\vec{M}| \tan(\Theta_0),\\
	\tan(\Theta_0)\approx|\vec{H}_{y1}|/|\vec{H}_0|
\end{align}
and a phase angle $\phi=\omega t$ where $\omega$ is the precession frequency.

\begin{figure}[ht]
\includegraphics[width=8cm]{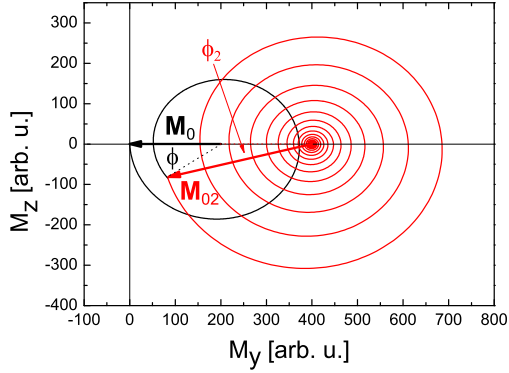}
\caption{Time evolution of $M_y$ and $M_z$ in a damped system excited by two subsequent pulses in +y-direction. Black curve: Precession from $t=0$ to $t=t_2$, Red curve: Precession after $t=t_2$.
\label{fig:plus_plus}}
\end{figure}

For coherent control a second field step $\vec{H}_{y2}$ is used at $t=t_2$ which can be applied either in +y or -y direction and with variable time delay. Depending on the direction of the second step this either tilts the magnetic field further away from $\vec{H}_{0}$ towards a new vector $\vec{H}_{2}$ or compensates $\vec{H}_{y1}$ and restores the external field $\vec{H}_{0}$.
Depending on the time delay and the sign of $\vec{H}_{y2}$ this affects the precession of $\vec{M}$ in different ways. At the time $t_2$ the magnetization starts to precess around a new axis with a new precession amplitude and a new phase angle.
For a detailed description we first consider the case of $\vec{H}_{y2}=\vec{H}_{y1}$. The center of the precession is then located at $y=2 |\vec{H}_{y2}|$ and $z=0$.
As we can see in figure \ref{fig:plus_plus} the new precession amplitude at $t=t_2$ which we name ${M}_{0_2}$ is given as

\begin{align}
{M}_{0_2}&=\sqrt{({M}_{0} [e^{-t_2/\tau} \sin{\omega t_2}])^2+({M}_{0} [1+e^{-t_2/\tau} \cos{\omega t_2}])^2}\\
&={M}_{0} (e^{-t_2/\tau})\sqrt{( \sin{\omega t_2})^2+(1/e^{-t_2/\tau} + \cos{\omega t_2})^2}
\end{align}

The precession angle $\phi_2$ at which the new precession starts can be calculated to

\begin{equation}
\phi_2=\arctan\left(\frac{\sin\omega t_2}{1/e^{-t_2/\tau}+\cos(\omega t_2)}\right)
\end{equation}

The time development of ${M}_y$ and ${M}_z$ can then be written as

\begin{align}
	{M}_y&= 2 {M}_{0} - {M}_{0_2} e^{-(t-t2)/\tau} \cos[\omega (t-t_2)+\phi_2]ßß\\
	{M}_z&=-{M}_{0_2} e^{-(t-t_2)/\tau} \sin[\omega (t-t_2)+\phi_2].
\end{align}

For a second field step of opposite direction the center of precession moves back to the original equilibrium direction $y=z=0$ and the previous equations are transformed to:

\begin{align}
	{M}_{0_2}&={M}_{0} (e^{-t_2/\tau})\sqrt{( \sin{\omega t_2})^2+(1/e^{-t_2/\tau} - \cos{\omega t_2})^2}\\
	\phi_2&=\arctan\left(\frac{-\sin\omega t_2}{1/e^{-t_2/\tau}-\cos(\omega t_2)}\right).
\end{align}

The time development of ${M}_y$ and ${M}_z$ can then be written as
\begin{align}
	{M}_y={M}_{0_2} e^{-(t-t_2)/\tau} \cos[\omega (t-t_2)+\phi_2]\\
	and \qquad {M}_z={M}_{0_2} e^{-(t-t_2)/\tau} \sin[\omega (t-t_2)+\phi_2].
\end{align}

In order to better visualize the consequences we first consider the case of no damping or $\tau=\infty$ and $\vec{H}_{y_1}$ in +y direction.
Given a precession period $1/\omega$  and $t\le t_2$ we have ${M}_z=0$  for all times $t=n\pi/\omega$ whenever n is an integer number corresponding to phase angles of integer multiples of $\pi$. For even n, we also have ${M}_{y}=0$. For uneven n we have ${M}_{y}=2{M}_{0}$. For even n a field step in +y direction results in the maximum of the new precession amplitude ${M}_{0_2}=2{M}_{0}$. The phase angle is $\phi_2=0$ so there is no phase jump. For uneven n a +y step has the opposite effect. The magnetization at that time is at $y=0$ and the field ${H}_2$ points along the x-axis. We thus have $\vec{M} \times \vec{H}\approx 0$ which effectively terminates the precession.
For -y pulses the situation is again reversed. Applying the second pulse at even n stops the precession while for uneven n the precession amplitude is maximized to ${M}_{0_2}=2{M}_{0}$.

For intermediate phase angles $\phi(t_2)$ we also obtain intermediate values for the precession amplitude. In general we can state that
a +y step reduces the precession amplitude for
\begin{equation}
(2n\pi+2\pi/3) < \phi(t_2) < (2n\pi+4\pi/3)
\end{equation}
while a -y step reduces the amplitude for
\begin{equation}
(2n\pi-\pi/3) < \phi(t_2) < (2n\pi+\pi/3).
\end{equation}
In the other cases the step in the respective direction leads to an increase of the amplitude. In all those cases a phase jump occurs ($\phi_2\ne 0$). The amplitude only remains unchanged for

\begin{equation}
\phi(t_2)=2n\pi\pm2\pi/3
\end{equation}

in case of a +y step and for

\begin{equation}
\phi(t_2)=2n\pi\pm\pi/3
\end{equation}

in case of a -y step. In those cases only the center of precession changes and a phase jump takes place.

We now include finite damping into our picture. For low damping the description given above is still a good approximation for small numbers of n meaning for the first (few) precession periods. For higher damping the picture becomes more complex, however, we can still evaluate a few special cases.
For $\phi_2=n\pi$ with even n and a +y step as well as for uneven n and a -y step the amplitude is still maximized, however, now the maximum amplitude is

\begin{equation}
{M}_{0_2}={M}_{0}(1+e^{-t_2/\tau})
\end{equation}

For even n and a -y step or for uneven n and a +y step the amplitude is minimized to

\begin{equation}
{M}_{0_2}={M}_{0}(1-e^{-t_2/\tau})
\end{equation}

but no longer reduced to zero.
Here a phase jump of $\pi$ occurs at $t=t_2$.

Obviously it is still possible to stop the precession completely if the height of the second field step $\vec{H}_{y2}$ is matched to $\vec{H}_{y2}=(e^{-t_2/\tau}-1)\vec{H}_{y1}$.

In the experimental setup the sample is placed face down on the signal line of a coplanar waveguide. Voltage steps are applied to the waveguide and induce the field steps $\vec{H}_{y1}$ and $\vec{H}_{y2}$. The resulting precession of the magnetization in the sample is detected by the induced voltage. A Sampling Oscilloscope (DSA $8300$) equipped with Time Domain Reflectometry (TDR) sampling modules ($80E08$ and $80E10$) is used to produce the voltage steps and to detect the induced signal. The voltage steps have amplitudes of $250$mV, a repetition rate of $200$kHz, and a respective rise time of $12-20$ps which depends on the used module. The polarity of the step and a delay time can be selected independently for each channel. The system can display the reflected and the transmitted signal as a function of time. For all time domain measurements a digital filter is used to suppress high frequency noise.
Because the voltage induced by the precession is extremely small compared to the applied step a reference measurement without precession must be subtracted as described by Silva et al. \cite{Silva1999}. A second set of coils is used to apply an external field in y-direction \cite{Silva1999}. In this configuration a field pulse in y direction does not induce any precession and yields the reference data which can be subtracted from the original measurement.

We investigate thin films of different ferromagnetic materials namely YIG, Permalloy and $La_{0.7}Sr_{0.3}MnO_{3}$ (LSMO). Here only the results for YIG are presented and discussed. However, it should be noted that for all investigated materials the results are in good agreement with micromagnetic simulations and frequency domain FMR experiments.

\begin{figure}[ht]
\includegraphics[width=8cm]{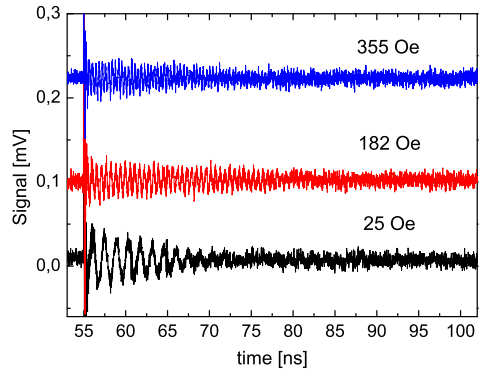}
\caption{Field dependent measurement of time-domain FMR.
\label{fig:H_dependence}}
\end{figure}

For the present experiment we use a $23nm$ thin YIG film fabricated by pulsed laser deposition (PLD) on a (111) oriented $Gd_{3}Ga_{5}O_{12}$ (GGG) substrate, using an oxygen pressure of $0.033$ mbar, a laser fluency of $2$ $J/cm^2$, laser repetition rate of $5$ Hz, a substrate temperature of $900 ^\circ$C and a growth rate of $0.5$ nm/min. In frequency domain FMR the film shows a linewidt of $8$ Oe at $10$GHz.
Figure \ref{fig:H_dependence} shows conventional time domain measurements at different fields $\vec{H}_0$ using only one voltage step to induce the precession of the magnetization. The high quality of the YIG film allows us to observe the oscillation over a time of $50ns$. At low magnetic fields a beating is observed (Fig. \ref{fig:H_dependence}, $25$Oe) which is caused by two resonances with slightly different frequencies ($0.68$GHz and $0.72$GHz) which can be determined by Fast Fourier Transformation (FFT). These multiple resonances are also confirmed by frequency domain FMR measurements.
In the following we apply two voltage steps with different time delays in order to investigate different cofigurations for coherent control. The two voltage steps are of opposite sign but similar magnitude corresponding to $\vec{H}_{y2}$ as a -y step which restores the original $\vec{H}_0$ after the second step.

\begin{figure}[ht]
\includegraphics[width=8cm]{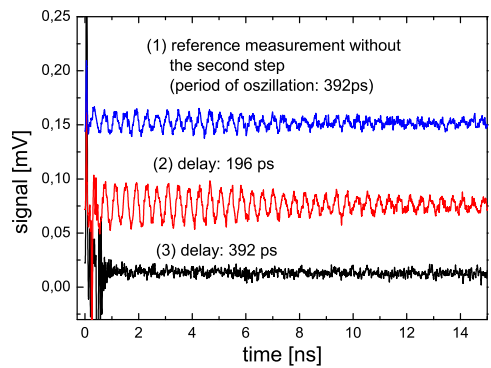}
\caption{Time-domain FMR, using one voltage step (1). With the second voltage step, applied with suitable delay times the precession can be maximized (2) or completely stopped (3).
\label{fig:step_min_max}}
\end{figure}

Figure \ref{fig:step_min_max} shows three examples, which are also theoretically discussed at the beginning of the paper. In the first case the precession is induced by one voltage step only to establish the ordinary precession pattern. In case $(2)$ the second voltage step is applied in the maximum of the oscillation. According to our theory this doubles the precession amplitude. Indeed we observe a signal which is approx. twice as large as for a single pulse. When the -y step is applied after a full precession corresponding to $\phi(t_2)=2\pi$ the precession is stopped (Fig. \ref{fig:step_min_max},(3)) as expected.
Besides the purely analytical approach to theory our experiments are also supported by micromagnetic simulations using OOMMF (\cite{oommf}).

\begin{figure}[ht]
\includegraphics[width=8cm]{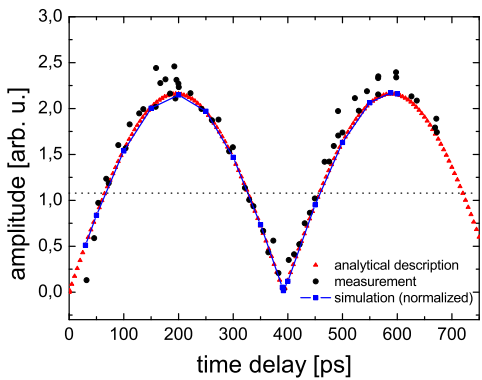}
\caption{Amplitudes of the precession versus the time delay. The dotted line shows the amplitude of the precession induced by only one voltage step and corresponds to the reference measurement in figure \ref{fig:step_min_max}. The simulated data have been normalized, assuming that the amplitudes of the oscillations induced by one voltage step are equal for the simulation and the experiment.
\label{fig:time_delay_amplitude}}
\end{figure}

Besides the two cases of $\phi(t_2)=\pi$ and $\phi(t_2)=2\pi$ also measurements for other delay times and angles are performed. In fig \ref{fig:time_delay_amplitude} the amplitude of the precession is plotted over the delay time. The diagram shows that by evaluating the time between the minima of the curve we can determine the procession period with very high accuracy, in fact much better than by using FFT or simply measuring the period of the oscillation in the measurement. Coherent control thus greatly enhances the precision of the measurement.

It should finally be noted that this method holds several advantages with respect to standard PIMM. By using a sequence of multiple alternating steps (in other words a sequence of pulses) we can massively enhance the precession amplitude for samples with low damping. As we know the precession amplitude after one period is reduced by a factor of $e^{-t/\tau}$. It is obvious that multiple excitations of identical magnitude which are applied after each period can increase the amplitude up to a magnitude ${M}_{max}$ which satisfies the condition:

\begin{equation}
{M}_{max}=\frac{M_{0}}{1- e^{-t/\tau}}
\end{equation}

This means that if the decay of the precession is only 1$\%$ a maximum gain of 100 is possible. In the presence of several resonances with only small difference in frequency using multiple steps or pulses can also help to facilitate the analysis. In PIMM two resonances close to each other result in a beating pattern in the time domain. Excitation with multiple steps which are spaced by just one period do little to suppress one of the two lines. If, however, the spacing is a suitable multiple of a precession period one of the resonances can be enhanced while the other one is suppressed. It may thus be possible to selectively excite a single resonance even if when its frequency is close to another resonance.

We have shown that by using two short magnetic field steps with varying time delay it is possible to coherently control ferromagnetic resonance in a PIMM geometry. The method can be used to increase the sensitivity of the method to enhance the temporal resolution and even to selectively excite single resonance modes even if other modes with different frequencies exist.

This work was funded by the DFG in the SFB 762 and by the EC in the project IFOX.


\end{document}